\newif\ifprlout
\prlouttrue

\ifprlout
\documentclass[twocolumn,showpacs,superscriptaddress,preprintnumbers,prl,aps]{revtex4}
\else
\documentclass[12pt,twoside,showpacs,letterpaper,doublespace]{article}
\fi

\usepackage{amssymb,amsmath}
\usepackage{graphicx}
\usepackage{dcolumn}
\usepackage{bm}

\usepackage{lineno}
\usepackage{setspace}

\newcommand{\babar}{{\mbox{\sl B\hspace{-0.4em} {\small\sl A}\hspace{-0.37em}\sl B\hspace{-0.4em}}} {\small\sl A\hspace{-0.02em}R}}

\newcommand{\ifb}{\ensuremath{\mathrm{fb^{-1}}}}
\newcommand{\TeV}{\ensuremath{\matqq	hrm{Te\kern -0.1em V}}}
\newcommand{\TeVc}{\ensuremath{\mathrm{Te\kern -0.1em V\!/}c}}
\newcommand{\TeVcc}{\ensuremath{\mathrm{Te\kern -0.1em V\!/}c^2}}
\newcommand{\GeV}{\ensuremath{\mathrm{Ge\kern -0.1em V}}}
\newcommand{\GeVc}{\ensuremath{\mathrm{Ge\kern -0.1em V\!/}c}}
\newcommand{\GeVcc}{\ensuremath{\mathrm{Ge\kern -0.1em V\!/}c^2}}
\newcommand{\MeV}{\ensuremath{\mathrm{Me\kern -0.1em V}}}
\newcommand{\MeVc}{\ensuremath{\mathrm{Me\kern -0.1em V\!/}c}}
\newcommand{\MeVcc}{\ensuremath{\mathrm{Me\kern -0.1em V\!/}c^2}}

\newcommand{\cdfii}{CDF\,II~}

\newcommand{\myto}{\kern -0.3em\to\kern -0.2em}
\newcommand{\Lxy}{L_{\rm xy}}
\newcommand{\BR}{{\mathcal {B}}}

\newcommand{\Bs}{\ensuremath{B^{0}_s}}

\newcommand{\phiphi}{\ensuremath{B_s^0 \myto \phi \phi}}
\newcommand{\psiphi}{\ensuremath{B_s^0 \myto J\!/\!\psi \phi}}

\newcommand{\psikstpdg}{\ensuremath{B^0 \myto J\!/\!\psi K^\ast(892)^0}}
\newcommand{\phikstpdg}{\ensuremath{B^0 \myto \phi K^\ast(892)^0}}

\newcommand{\jpsi}{\ensuremath{J\!/\!\psi}}

\newcommand{\mpmm}{\ensuremath{\mu^+\mu^-}}
\newcommand{\KKpm}{\ensuremath{K^+K^-}}

\newcommand{\CP}{\ensuremath{C\!P}}

\newcommand{\T}{\ensuremath{T}}

\newcommand{\BVV}{\ensuremath{B\myto VV}}

\newcommand{\phiKK}{\ensuremath{\phi \myto \KKpm}}
\newcommand{\psimm}{\ensuremath{J\!/\!\psi \myto \mpmm}}

\newcommand{\fL}{\ensuremath{f_{\textup{L}}\ }}

\newcommand{\tp}{\ensuremath{\mathrm{TP}}}

\newcommand{\tpu}{\ensuremath{\mathrm{TP_1}}}
\newcommand{\tpd}{\ensuremath{\mathrm{TP_2}}}
\newcommand{\atp}{\ensuremath{\mathcal{A}_{\mathrm{TP}}}}
\newcommand{\atpu}{\ensuremath{\mathcal{A}^1_{\mathrm{TP}}}}
\newcommand{\atpd}{\ensuremath{\mathcal{A}^2_{\mathrm{TP}}}}

\newcommand{\coscos}{\ensuremath{\cos\vartheta_1\cos\vartheta_2}}
\newcommand{\stat}{\ensuremath{\text{(stat)}}}
\newcommand{\syst}{\ensuremath{\text{(syst)}}}

\begin{document}

\ifprlout
\else
\pagewiselinenumbers
\doublespace
\fi

\ifprlout

\title{
{
\vspace*{-0.9cm}
\raggedleft\small
{\tt FERMILAB-PUB-11-345-E\\
CDF/PUB/BOTTOM/PUBLIC/10383}\\
\hfill\\
}
Measurement of Polarization and Search for \CP\ Violation in \phiphi\ Decays} 

\affiliation{Institute of Physics, Academia Sinica, Taipei, Taiwan 11529, Republic of China} 
\affiliation{Argonne National Laboratory, Argonne, Illinois 60439, USA} 
\affiliation{University of Athens, 157 71 Athens, Greece} 
\affiliation{Institut de Fisica d'Altes Energies, ICREA, Universitat Autonoma de Barcelona, E-08193, Bellaterra (Barcelona), Spain} 
\affiliation{Baylor University, Waco, Texas 76798, USA} 
\affiliation{Istituto Nazionale di Fisica Nucleare Bologna, $^{aa}$University of Bologna, I-40127 Bologna, Italy} 
\affiliation{University of California, Davis, Davis, California 95616, USA} 
\affiliation{University of California, Los Angeles, Los Angeles, California 90024, USA} 
\affiliation{Instituto de Fisica de Cantabria, CSIC-University of Cantabria, 39005 Santander, Spain} 
\affiliation{Carnegie Mellon University, Pittsburgh, Pennsylvania 15213, USA} 
\affiliation{Enrico Fermi Institute, University of Chicago, Chicago, Illinois 60637, USA}
\affiliation{Comenius University, 842 48 Bratislava, Slovakia; Institute of Experimental Physics, 040 01 Kosice, Slovakia} 
\affiliation{Joint Institute for Nuclear Research, RU-141980 Dubna, Russia} 
\affiliation{Duke University, Durham, North Carolina 27708, USA} 
\affiliation{Fermi National Accelerator Laboratory, Batavia, Illinois 60510, USA} 
\affiliation{University of Florida, Gainesville, Florida 32611, USA} 
\affiliation{Laboratori Nazionali di Frascati, Istituto Nazionale di Fisica Nucleare, I-00044 Frascati, Italy} 
\affiliation{University of Geneva, CH-1211 Geneva 4, Switzerland} 
\affiliation{Glasgow University, Glasgow G12 8QQ, United Kingdom} 
\affiliation{Harvard University, Cambridge, Massachusetts 02138, USA} 
\affiliation{Division of High Energy Physics, Department of Physics, University of Helsinki and Helsinki Institute of Physics, FIN-00014, Helsinki, Finland} 
\affiliation{University of Illinois, Urbana, Illinois 61801, USA} 
\affiliation{The Johns Hopkins University, Baltimore, Maryland 21218, USA} 
\affiliation{Institut f\"{u}r Experimentelle Kernphysik, Karlsruhe Institute of Technology, D-76131 Karlsruhe, Germany} 
\affiliation{Center for High Energy Physics: Kyungpook National University, Daegu 702-701, Korea; Seoul National University, Seoul 151-742, Korea; Sungkyunkwan University, Suwon 440-746, Korea; Korea Institute of Science and Technology Information, Daejeon 305-806, Korea; Chonnam National University, Gwangju 500-757, Korea; Chonbuk National University, Jeonju 561-756, Korea} 
\affiliation{Ernest Orlando Lawrence Berkeley National Laboratory, Berkeley, California 94720, USA} 
\affiliation{University of Liverpool, Liverpool L69 7ZE, United Kingdom} 
\affiliation{University College London, London WC1E 6BT, United Kingdom} 
\affiliation{Centro de Investigaciones Energeticas Medioambientales y Tecnologicas, E-28040 Madrid, Spain} 
\affiliation{Massachusetts Institute of Technology, Cambridge, Massachusetts 02139, USA} 
\affiliation{Institute of Particle Physics: McGill University, Montr\'{e}al, Qu\'{e}bec, Canada H3A~2T8; Simon Fraser University, Burnaby, British Columbia, Canada V5A~1S6; University of Toronto, Toronto, Ontario, Canada M5S~1A7; and TRIUMF, Vancouver, British Columbia, Canada V6T~2A3} 
\affiliation{University of Michigan, Ann Arbor, Michigan 48109, USA} 
\affiliation{Michigan State University, East Lansing, Michigan 48824, USA}
\affiliation{Institution for Theoretical and Experimental Physics, ITEP, Moscow 117259, Russia}
\affiliation{University of New Mexico, Albuquerque, New Mexico 87131, USA} 
\affiliation{Northwestern University, Evanston, Illinois 60208, USA} 
\affiliation{The Ohio State University, Columbus, Ohio 43210, USA} 
\affiliation{Okayama University, Okayama 700-8530, Japan} 
\affiliation{Osaka City University, Osaka 588, Japan} 
\affiliation{University of Oxford, Oxford OX1 3RH, United Kingdom} 
\affiliation{Istituto Nazionale di Fisica Nucleare, Sezione di Padova-Trento, $^{bb}$University of Padova, I-35131 Padova, Italy} 
\affiliation{LPNHE, Universite Pierre et Marie Curie/IN2P3-CNRS, UMR7585, Paris, F-75252 France} 
\affiliation{University of Pennsylvania, Philadelphia, Pennsylvania 19104, USA}
\affiliation{Istituto Nazionale di Fisica Nucleare Pisa, $^{cc}$University of Pisa, $^{dd}$University of Siena and $^{ee}$Scuola Normale Superiore, I-56127 Pisa, Italy} 
\affiliation{University of Pittsburgh, Pittsburgh, Pennsylvania 15260, USA} 
\affiliation{Purdue University, West Lafayette, Indiana 47907, USA} 
\affiliation{University of Rochester, Rochester, New York 14627, USA} 
\affiliation{The Rockefeller University, New York, New York 10065, USA} 
\affiliation{Istituto Nazionale di Fisica Nucleare, Sezione di Roma 1, $^{ff}$Sapienza Universit\`{a} di Roma, I-00185 Roma, Italy} 

\affiliation{Rutgers University, Piscataway, New Jersey 08855, USA} 
\affiliation{Texas A\&M University, College Station, Texas 77843, USA} 
\affiliation{Istituto Nazionale di Fisica Nucleare Trieste/Udine, I-34100 Trieste, $^{gg}$University of Udine, I-33100 Udine, Italy} 
\affiliation{University of Tsukuba, Tsukuba, Ibaraki 305, Japan} 
\affiliation{Tufts University, Medford, Massachusetts 02155, USA} 
\affiliation{University of Virginia, Charlottesville, Virginia 22906, USA}
\affiliation{Waseda University, Tokyo 169, Japan} 
\affiliation{Wayne State University, Detroit, Michigan 48201, USA} 
\affiliation{University of Wisconsin, Madison, Wisconsin 53706, USA} 
\affiliation{Yale University, New Haven, Connecticut 06520, USA} 
\author{T.~Aaltonen}
\affiliation{Division of High Energy Physics, Department of Physics, University of Helsinki and Helsinki Institute of Physics, FIN-00014, Helsinki, Finland}
\author{B.~\'{A}lvarez~Gonz\'{a}lez$^w$}
\affiliation{Instituto de Fisica de Cantabria, CSIC-University of Cantabria, 39005 Santander, Spain}
\author{S.~Amerio}
\affiliation{Istituto Nazionale di Fisica Nucleare, Sezione di Padova-Trento, $^{bb}$University of Padova, I-35131 Padova, Italy} 

\author{D.~Amidei}
\affiliation{University of Michigan, Ann Arbor, Michigan 48109, USA}
\author{A.~Anastassov}
\affiliation{Northwestern University, Evanston, Illinois 60208, USA}
\author{A.~Annovi}
\affiliation{Laboratori Nazionali di Frascati, Istituto Nazionale di Fisica Nucleare, I-00044 Frascati, Italy}
\author{J.~Antos}
\affiliation{Comenius University, 842 48 Bratislava, Slovakia; Institute of Experimental Physics, 040 01 Kosice, Slovakia}
\author{G.~Apollinari}
\affiliation{Fermi National Accelerator Laboratory, Batavia, Illinois 60510, USA}
\author{J.A.~Appel}
\affiliation{Fermi National Accelerator Laboratory, Batavia, Illinois 60510, USA}
\author{A.~Apresyan}
\affiliation{Purdue University, West Lafayette, Indiana 47907, USA}
% ...
\collaboration{CDF Collaboration\footnote{With visitors from $^a$Istituto Nazionale di Fisica Nucleare, Sezione di Cagliari, 09042 Monserrato (Cagliari), Italy,
$^b$University of CA Irvine, Irvine, CA  92697, USA,
$^c$University of CA Santa Barbara, Santa Barbara, CA 93106, USA,
$^d$University of CA Santa Cruz, Santa Cruz, CA  95064, USA,
$^e$CERN,CH-1211 Geneva, Switzerland,
$^f$Cornell University, Ithaca, NY  14853, USA, 
$^g$University of Cyprus, Nicosia CY-1678, Cyprus, 
$^h$Office of Science, U.S. Department of Energy, Washington, DC 20585, USA,
$^i$University College Dublin, Dublin 4, Ireland,
$^j$University of Fukui, Fukui City, Fukui Prefecture, Japan 910-0017,
$^k$Universidad Iberoamericana, Mexico D.F., Mexico,
$^l$Iowa State University, Ames, IA  50011, USA,
$^m$University of Iowa, Iowa City, IA  52242, USA,
$^n$Kinki University, Higashi-Osaka City, Japan 577-8502,
$^o$Kansas State University, Manhattan, KS 66506, USA,
$^p$University of Manchester, Manchester M13 9PL, United Kingdom,
$^q$Queen Mary, University of London, London, E1 4NS, United Kingdom,
$^r$University of Melbourne, Victoria 3010, Australia,
$^s$Muons, Inc., Batavia, IL 60510, USA,
$^t$Nagasaki Institute of Applied Science, Nagasaki, Japan, 
$^u$National Research Nuclear University, Moscow, Russia,
$^v$University of Notre Dame, Notre Dame, IN 46556, USA,
$^w$Universidad de Oviedo, E-33007 Oviedo, Spain, 
$^x$Texas Tech University, Lubbock, TX  79609, USA,
$^y$Universidad Tecnica Federico Santa Maria, 110v Valparaiso, Chile,
$^z$Yarmouk University, Irbid 211-63, Jordan,
$^{hh}$On leave from J.~Stefan Institute, Ljubljana, Slovenia, 
}}
\noaffiliation


\else

\begin{center}
\begin{Large}
Measurement of Polarization and Search for \CP\ Violation in \phiphi\ Decays
\end{Large}
\end{center}

\vspace{1in}
\begin{center}
\begin{large}{The CDF Collaboration\\}
\end{large}
\today
\end{center}

\fi

\date{\today}

\begin{abstract}

We present the first measurement of polarization and \CP--violating asymmetries in a 
\Bs\ decay into two light vector mesons, \phiphi, and an improved determination
of its branching ratio using $295$ decays reconstructed in a data sample corresponding to 2.9
\ifb\ of integrated luminosity collected by the CDF experiment at the Fermilab Tevatron collider. 
The fraction of longitudinal polarization  is determined to be 
$\fL = 0.348 \pm 0.041 ({\rm stat}) \pm 0.021 ({\rm syst})$,
and the branching ratio $\BR (\phiphi) = [2.32 \pm 0.18 ({\rm stat}) \pm 0.82 ({\rm syst})] \times 10^{-5}$.
Asymmetries of decay angle distributions sensitive to \CP\ violation are
measured to be
$  A_u = - 0.007 \pm 0.064 \stat \pm 0.018 \syst$ and $ A_v = -0.120 \pm 0.064 \stat \pm 0.016\syst$.

\end{abstract}

\ifprlout
\maketitle
\fi

%
%
 Several charmless \Bs\ decays were observed at the Tevatron in Run II~\cite{prlphix,BsKK}, 
 but a detailed investigation of  decay properties and of \CP\ violation in these decays is still lacking.
 The $\phiphi$ process is mediated  by a one--loop flavor--changing neutral current, 
 the $b \myto s$ penguin, and belongs to the class of decays  where the final state
 consists of a pair of light spin--1 mesons (V).
 Three independent amplitudes govern \BVV\ decays, 
 corresponding to the polarizations of the final--state
 vector mesons: longitudinal polarization, and transverse
 polarization with spins parallel or perpendicular to each other. 
 The first two states are \CP--even, while the
 last one is \CP--odd. Polarization amplitudes can be measured analyzing 
 angular distributions of final--state particles. Interference between 
 the \CP--even and \CP--odd amplitudes can generate asymmetries
 in angular distributions, the triple product (TP) asymmetries, which may signal unexpected  \CP\ violation  
 due to physics beyond the standard model (SM).
\newline
\indent
  The V--A structure of charged weak currents leads to the 
 expectation of a dominant longitudinal polarization~\cite{BVV-QCDf,Ali:2007ff}. 
 Approximately equal longitudinal and transverse polarizations  
 have been measured instead in $b\myto s$ penguin--dominated 
 $B^0$ and $B^+$ decay modes~\cite{phikstar}. 
 This is explained in the SM by including either non--factorizable 
 penguin--annihilation effects~\cite{puzzle-PA} 
 or final state interactions~\cite{puzzle-FSI}. 
 Recent theoretical predictions~\cite{BVV-QCDf,Ali:2007ff} indicate a longitudinal fraction $f_L$ in the 40--70 \% range,
 when phenomenological parameters are adjusted to accommodate present experimental data.
 Explanations involving new physics (NP) in the $b\myto s$ 
 penguin process have also been proposed~\cite{NPtheory}.
 Additional experimental information in \Bs\ penguin--dominated decays, 
 such as \phiphi, may help distinguishing the various solutions~\cite{Datta:2008wf},
 and can be used to derive upper limits for the mixing--induced \CP\ asymmetries~\cite{Bartsch:2008ps}.
\newline
\indent
%
%
%
%
%
%
Triple product asymmetries are odd under time-reversal (\T), and can be generated
either by final--state interactions or \CP\ violation.  
In flavor--untagged samples, where the initial $B$ flavor 
is not identified, TP asymmetries can be shown to
signify genuine \CP\ violation~\cite{TPtruefake}.  In this respect they are very sensitive
to the presence of NP in the decay since they do not require a 
strong--phase difference between NP and SM amplitudes, as opposed to
direct \CP\ asymmetries~\cite{TPtheory}.
%
The TP asymmetry is defined as
$
\atp=\frac{\Gamma(\tp>0)-\Gamma(\tp<0)}
{\Gamma(\tp>0)+\Gamma(\tp<0)},
$
where $\Gamma$ is the decay width for the given process.
In \phiphi\ decays two TP asymmetries can be studied, corresponding to the two interference terms
between amplitudes with different \CP. These asymmetries are predicted to vanish in the SM, and 
an observation of a non--zero asymmetry would be an unambigous
sign of NP~\cite{TPtheory}.\newline
\indent
 In this Letter we present the first measurement of polarization amplitudes and of TP asymmetries in the \phiphi\ decay and an updated measurement of its branching ratio 
 using \psiphi\ decays reconstructed in the same dataset as a normalization.
 Data from an integrated luminosity of 2.9 \ifb\ of  $p\bar p$~collisions  at $\sqrt{s}=1.96$~TeV are analyzed.
\newline
\indent 
The components of the \cdfii detector relevant for this analysis are briefly 
described below; a more complete description can be found elsewhere~\cite{cdf}. 
We reconstruct charged--particle trajectories (tracks) in the pseudorapidity 
range $|\eta|\lesssim 1$~\cite{CDFsystem} using a silicon microstrip 
vertex detector~\cite{vertex-detectors} 
and a central drift chamber~\cite{cot}, 
both immersed in a $1.4\,{\rm T}$ solenoidal magnetic field. 
%
%
The detection of muons in the pseudorapidity
range $|\eta|\lesssim 0.6$ is provided by two sets of drift 
chambers located behind the calorimeters (CMU) and behind additional
steel absorbers (CMP), while the CMX detector covers the range 
$0.6 \lesssim |\eta|\lesssim 1.0$~\cite{cmu}. 
%
%
A sample enriched with heavy--flavor particles is 
selected by the displaced--track trigger~\cite{trigger}, based on
the Silicon Vertex Trigger (SVT)~\cite{svt}. It
provides a precise measurement of the track impact parameter ($d_0$), 
defined as the distance of closest approach to the beam axis in the transverse plane.
Decays of heavy--flavor particles are identified by requiring 
two tracks with $120\,\mu{\rm m}\le d_0 \le1.0\rm\,mm$ and 
applying a requirement on the two--dimensional decay length, 
$\Lxy >200\rm\,\mu m$~\cite{Lxy}.
\newline
\indent We reconstruct \Bs\ mesons  by first forming  \phiKK\ and \psimm\
candidate decays from opposite--sign track pairs 
with mass within 15 and 100 \MeVcc\ of the known~\cite{pdg2010} $\phi$ and \jpsi\ mass, respectively.
At least one \jpsi\ track  is required to match 
a segment reconstructed in the muon detectors.
We form \phiphi\ (\psiphi) candidates by fitting to a single vertex
the $\phi$ $\phi$ ($\jpsi$ $\phi$) candidate pairs.
In the \psiphi\ case the fit
constrains the  mass of the two
muons  to the \jpsi\ mass~\cite{pdg2010}.
At least one pair of tracks in the \Bs\ candidate must satisfy  
the trigger requirements. 
%
%
 Combinatorial background and partially reconstructed decays are reduced by exploiting the 
long lifetime and relatively hard $p_T$ spectrum of \Bs\ mesons. 
We follow closely the selection adopted in~\cite{prlphix}, 
using the  vertex fit $\chi^2$, the $\Lxy$, the reconstructed
\Bs\ and $\phi$ meson impact parameters, and the minimum kaon transverse
momentum as discriminating variables. 
The selection requirements are set by maximizing the quantity
$\mathrm {S/\sqrt{S+B}}$, where the accepted
number of signal events $\mathrm{S}$ is derived from a
Monte Carlo (MC) simulation~\cite{cdfsim} of the \cdfii detector and trigger,
while the number of background events $\mathrm{B}$ is modeled using data in mass
sideband regions:
$(5.02,5.22)$ and $(5.52,5.72)$ \GeVcc. 
The resulting mass distributions are shown in Fig.~\ref{fig_peaks}.\newline
\begin{figure} 
\includegraphics[width=0.5\linewidth]{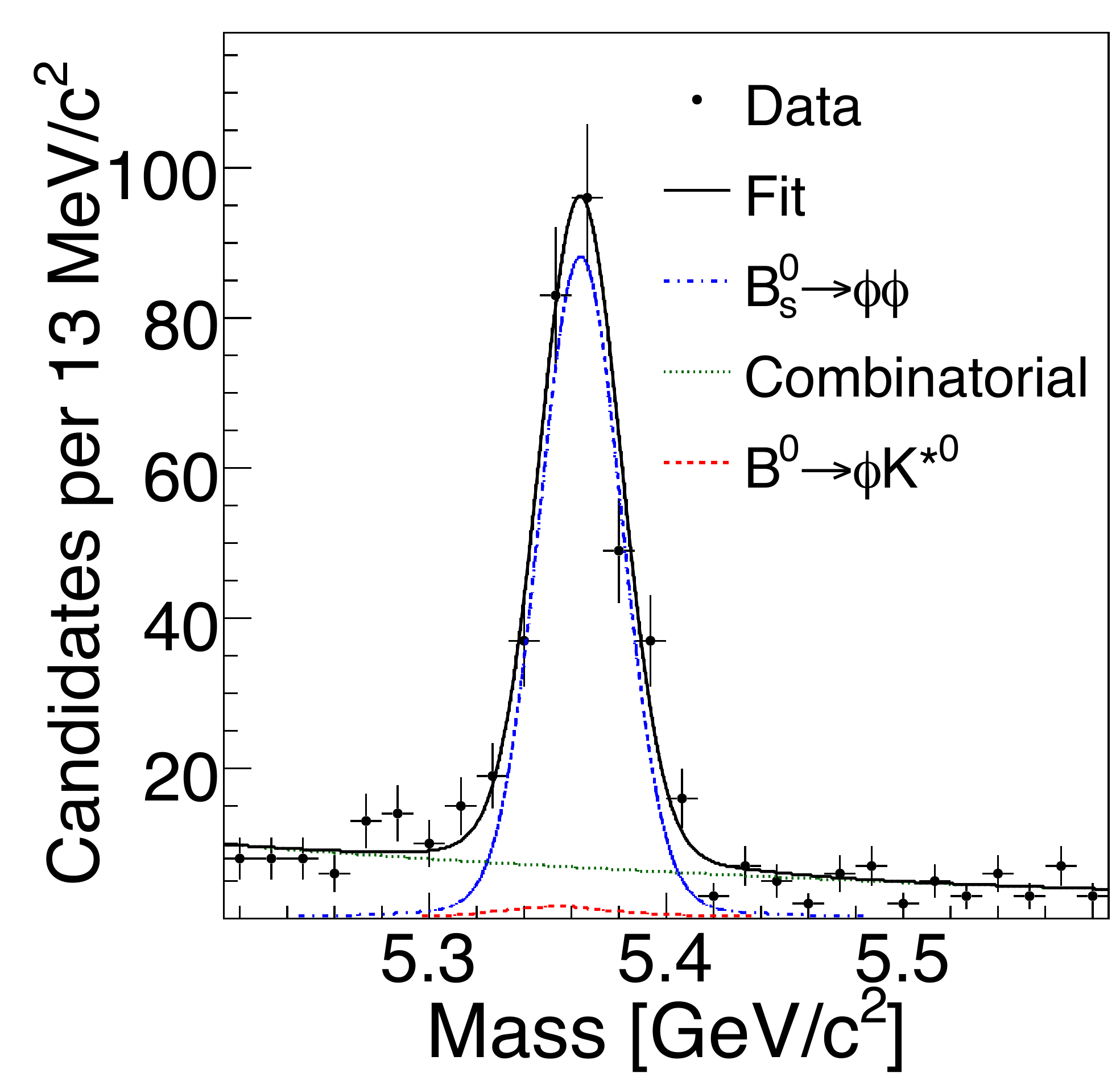}\hfill
\includegraphics[width=0.5\linewidth]{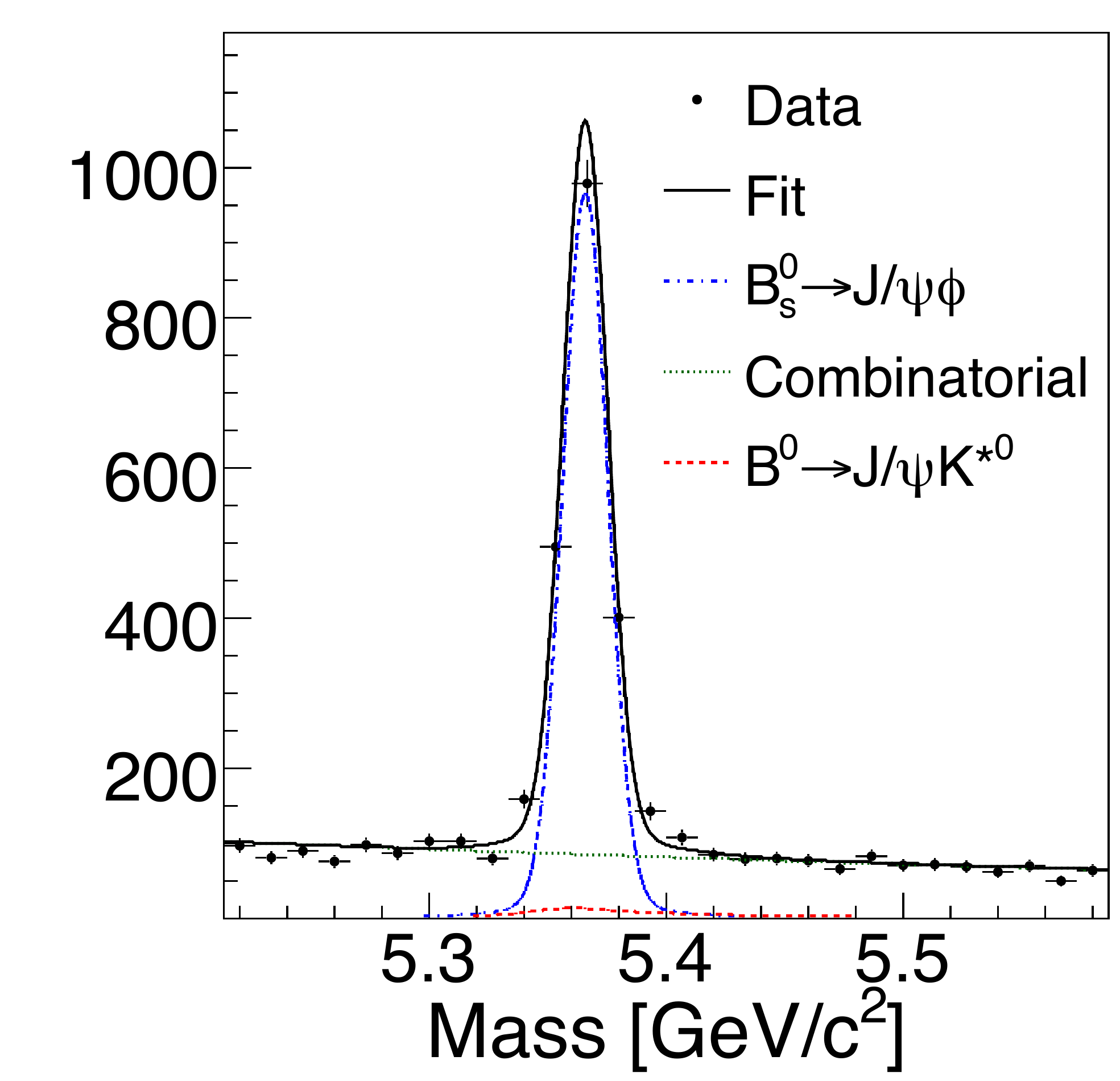} 
\caption{\label{fig_peaks}
The invariant mass of the four kaons (left) and of  the \jpsi\ and two
kaons (right) for \phiphi\ and \psiphi\ candidates,
overlayed with fit projections and separate signal and background components.
The narrower signal peak for the $\psiphi$ 
is due to the $\jpsi$ mass constraint applied in the reconstruction.
%
\vspace{-0.3cm}}
\end{figure}
\indent
A binned maximum likelihood (ML) fit to the $m_B$  distribution is performed to determine
the \Bs\ yield for both decay modes.  The signal is parameterized by two Gaussian functions 
with the same mean value, but different widths. The ratios 
between the two widths and between the integrals of the two components are fixed
based on MC simulations.  
The combinatorial background has a smooth mass             
distribution near the signal and is modeled with an exponential function.
A reflection from \phikstpdg\ (\psikstpdg) with misassigned kaon mass to final state pions contaminates the \phiphi\ (\psiphi) signal region. 
Parameterizations and efficiencies determined from simulation are used for these backgrounds.
Their normalizations are derived from the known~\cite{pdg2010} branching 
ratios,  fragmentation fraction ratio $f_s/f_d$, and the ratio of the detection efficiencies 
relative to signal ones. 
We estimate  $(4.19 \pm 0.93) \%$ and $(2.7 \pm 1.0)\%$ 
reflection background under the \psiphi\ and \phiphi\ signals, respectively.
Free parameters of the fit are the signal fraction, the \Bs\ mass $M$, and 
width $\sigma$, together with
the exponential slope $b_0$ defining the combinatorial background mass shape.
We estimate the total number of signal decays as $N_{\phi \phi} = 295 \pm 20 ({\rm stat}) \pm 12  ({\rm syst})$  
and $N_{\psi \phi} = 1766 \pm 48 ({\rm stat}) \pm 41 ({\rm syst})$, where the systematic uncertainty 
is estimated by varying signal and background models.\newline
%
\indent
 The \phiphi\ decay rate is derived from the relation
\begin{eqnarray*}
\label{eq:BRphiphi}
\frac{\BR\left(\phiphi\right)}
     {\BR\left(\psiphi\right)} =
\frac{N_{\phi\phi}}{N_{\psi\phi}}
\frac{\BR\left(J/\psi\to\mu\mu\right)}{\BR \left(\phi\to K K \right)}
\frac{\epsilon_{\psi\phi}}{\epsilon_{\phi\phi}}\, \epsilon^{\mu}_{\psi\phi}\ ,
\end{eqnarray*}
%
%
where  $\epsilon_{\psi\phi}/\epsilon_{\phi\phi}$ is the
acceptance times efficiency ratio for the two decays and
$\epsilon^{\mu}_{\psi\phi}$ is the efficiency for identifying 
at least one of the two muons.
The efficiency ratio is determined using a MC simulation of the \cdfii detector and trigger,
whose reliability  in determining relative trigger and 
reconstruction efficiencies has been verified for
several different decay modes also using data--driven approaches~\cite{earlierCDFBR}.
We estimate $\epsilon_{\psi\phi}/\epsilon_{\phi\phi}=0.939 \pm 0.099$,  
where the uncertainty includes systematic effects from polarization uncertainties 
in the two decay modes (9\%), from the different trigger efficiencies for kaons and muons (4\%), and 
from the \Bs\ $p_{T}$ spectra (1\%).
%
 We use inclusive \jpsi\ data to derive the single--muon identification efficiency 
as a function of muon $p_{T}$.
It is determined separately in two pseudorapidity regions corresponding, respectively, 
to the CMU/CMP and CMX detectors,
and is described by a turn-on function that depends on a 
plateau, a slope, and a threshold parameter.
We use simulated \psiphi\ decays to calculate
$\epsilon^{\mu}_{\psi\phi}$ treating the efficiencies 
for the two muons as uncorrelated:
$ \epsilon^{\mu}_{\psi\phi} = (86.95 \pm 0.44 ({\rm stat}) \pm 0.75 ({\rm syst}))\%$. 
The systematic uncertainty includes the uncertainty on the background
subtraction and effects of residual correlation between the two muon efficiencies.\newline
%
%
\indent 
We measure $\BR (\phiphi)/\BR(\psiphi) = [1.78 \pm 0.14 ({\rm stat}) \pm 0.20 ({\rm syst})] \times 10 ^ {-2} $ and
derive $ \BR (\phiphi) = [2.32 \pm 0.18 ({\rm stat}) \pm 0.26 ({\rm syst}) \pm 0.78 ({\rm br})] \times 10 ^{-5} $,
using the known~\cite{pdg2010} $\BR(\psiphi)$, which contributes the dominant uncertainty, 
labeled (br).
%
%
This result is in agreement and supersedes our
previous measurement~\cite{prlphix} with
a substantial reduction of its statistical uncertainty; it is also
consistent with recent theoretical calculations~\cite{BVV-QCDf,Ali:2007ff}.\newline
%
%
%
\indent
  We describe the angular distribution of the \phiphi\ decay products using the helicity variables
$\vec{\omega}=(\cos\vartheta_1,\cos\vartheta_2,\varPhi$), where
$\vartheta_i$ is the angle between the direction of the $K^+$ from each
$\phi$ and the direction opposite the $B^0_s$ in the vector
meson rest frame, and $\varPhi$ is the angle between the two resonance
decay planes in the $B^0_s$ rest frame. The three independent complex amplitudes are
$A_0$ for the longitudinal polarization and $A_\parallel$ ($A_\perp$) 
for transverse polarization with spins parallel (perpendicular) to each other.
They are related by $|A_0|^2+|A_\parallel|^2+|A_\perp|^2=1$.
The differential decay rate is expressed as 
$d^4\Gamma/(dt d\vec{\omega})  \propto \sum_{i=1}^6 K_i(t)f_i(\vec{\omega})$,
where  the functions $K_i(t)$ encode the \Bs\ time evolution including mixing and depend 
on the polarization amplitudes,
and the $f_i(\vec{\omega})$ are functions of the helicity
angles only~\cite{TPtheory}. To extract the polarization amplitudes we measure the time-integrated 
angular distribution assuming no direct \CP\ violation 
and a negligible weak phase difference
between \Bs\ mixing and \phiphi\ decay as predicted in the SM.
The time--integrated differential decay rate
depends on the polarization amplitudes at $t=0$ and on
the light and heavy $B_s^0$ mass--eigenstate lifetimes, $\tau_{\text{L}}$ and $\tau_{\text{H}}$,
as follows:
\begin{eqnarray}
\label{eq:decay_rate}
\frac{d^3\Gamma}{d\vec{\omega}} &\propto &\, \tau _{\textup{L}}\big(
|A_0|^2f_1 (\vec{\omega}) + |A_\parallel|^2 f_2 (\vec{\omega}) \nonumber \\ 
& &+ |A_0||A_\parallel|\cos\delta_\parallel f_5 (\vec{\omega})\big) + \tau_{\textup{H}}|A_\perp|^2f_3 (\vec{\omega}),
\end{eqnarray}
where $\delta_\parallel=\arg (A_0^\star A_\parallel)$ and
\begin{equation*}
\begin{split}
f_1(\vec{\omega})&=4\cos^2\vartheta_1\cos^2\vartheta_2, \\
f_2(\vec{\omega})&=\sin^2\vartheta_1\sin^2\vartheta_2(1+\cos 2\varPhi),\\
f_3(\vec{\omega})&=\sin^2\vartheta_1\sin^2\vartheta_2(1-\cos 2\varPhi),\\
f_5(\vec{\omega})&=\sqrt{2}\sin 2\vartheta_1 \sin 2\vartheta_2 \cos\varPhi.
\end{split}
\end{equation*}
%
%
Two triple products are present in \BVV\ decays: $\tpd \equiv  \Im(A_\parallel^\star A_\perp) $,
and  $\tpu \equiv \Im(A_0^\star A_\perp)$. These factors appear, respectively, in the 
decay rate terms $K_4(t)$ and $K_6(t)$ multiplied by the functions
\begin{equation*}
\begin{split}
f_4(\vec{\omega})&=-2\sin^2\vartheta_1 \sin^2\vartheta_2\sin 2\varPhi,\\
f_6(\vec{\omega})&=-\sqrt{2}\sin 2\vartheta_1 \sin 2\vartheta_2 \sin\varPhi.
\end{split}
\end{equation*}
In flavor--untagged samples the TP terms, that vanish in the absence of NP, are proportional to the so--called 
\emph{true} triple products, and provide two \CP--violating observables, \atpu\ and \atpd~\cite{TPtruefake}. 
We access \atpd\ through the observable
$u=\sin 2\varPhi$. We measure the $u$ asymmetry, $A_u$, by integrating over $\cos \vartheta_{1,2}$
the untagged decay rate and counting events with  $u\!>\!0$ ($N^+_u$)
and $u\!<\!0$ ($N^-_u$).
Similarly, \atpu\ is accessed through an asymmetry in $\sin\varPhi$. 
We define the observable $v$  
as $v=\sin\varPhi$ ($v=-\sin\varPhi$) if  $\coscos\geq 0$ ($\coscos<0$)
and measure its asymmetry $A_v$ by counting events with $v\!>\!0$ ($N^+_v$)
and $v\!<\!0$ ($N^-_v$). The asymmetries are defined as
\begin{eqnarray}
\label{eq:atp2}
& A_{u(v)} &= \frac{N^+_{u(v)} - N^-_{u(v)}}{N^+_{u(v)} + N^-_{u(v)}} = \mathcal{N}_{u(v)} \times \nonumber\\ 
& & \big[\Im(A^\star_{\parallel(0)}A_\perp) 
+ \Im(\bar{A}^\star_{\parallel(0)} \bar{A}_\perp)\big]=\mathcal{N}_{u(v)}   
\mathcal{A}^{2(1)}_{\rm{TP}},
\end{eqnarray}
where the two normalization factors are $\mathcal{N}_u=-2/\pi$ and
 $\mathcal{N}_v=-\sqrt{2}/\pi$. 
Both $A_u$ and $A_v$ are proportional to \CP--violating TP asymmetries, 
and are also sensitive to mixing--induced TP when considering
the decay--width difference of the \Bs\ system.\newline
%
\indent
We perform an unbinned ML fit to the reconstructed
mass of the \Bs\ candidates and the helicity
angles in order to measure the polarization amplitudes.
%
The contribution of each candidate to the likelihood is
$\mathcal{L}_i=f_s\mathcal{P}_s(m_{B\,i},\vec{\omega}_i|\vec{\xi}_s)+(1-f_s)\mathcal{P}_b(m_{B\,i},\vec{\omega}_i|\vec{\xi}_b)$,
where $f_s$ is the signal fraction and
$\mathcal{P}_j$ are the probability density functions
(PDFs) for the \phiphi\ signal ($j=s$) and background ($j=b$) components, which
depend on the fit parameters $\vec{\xi}_s$ and $\vec{\xi}_b$,
respectively.  The effects of neglecting the reflection background are included in  
the systematic uncertainties.
Both the signal and the background PDFs are the products of a mass component, described 
earlier, and an angular one.
%
The signal angular component is given by Eq.~\ref{eq:decay_rate} multiplied by
an acceptance factor. The acceptance is computed in
bins of the helicity angles from simulated \phiphi\ decays averaged over all possible
spin states of the decay products and passed through detector simulation,
full reconstruction, and analysis cuts. 
%
%
%
We use an empirical parameterization derived from the observed angular
distributions in the mass sidebands to model the background angular
PDF: the product of a flat distribution for the $\varPhi$ angle and a parabolic
function for the other two, whose single parameter $b_1$ is a fit parameter.
We fix $\tau_{\textup{L}}$ and $\tau_{\textup{H}}$ to the world
average values~\cite{pdg2010}. There are eight free parameters in the fit: $f_s$, 
$\vec{\xi}_s=(M,\sigma,|A_0|^2,|A_\parallel|^2,\cos\delta_\parallel)$
and $\vec{\xi}_b=(b_0,b_1)$.
%
%
The fit has been extensively tested using simulated
samples with a variety of input parameters 
and shows unbiased estimates of parameters and their uncertainties.
We also perform the polarization measurement using the sample of $\approx$1700 \psiphi\ candidates 
described earlier. We find $|A_0|^2=0.534\pm0.019 ({\rm stat})$ and $|A_\parallel|^2=0.220\pm0.025 ({\rm stat})$, 
in good agreement with current measurements~\cite{psiphiparam}. 
\begin{figure} 
\includegraphics[width=1.\linewidth]{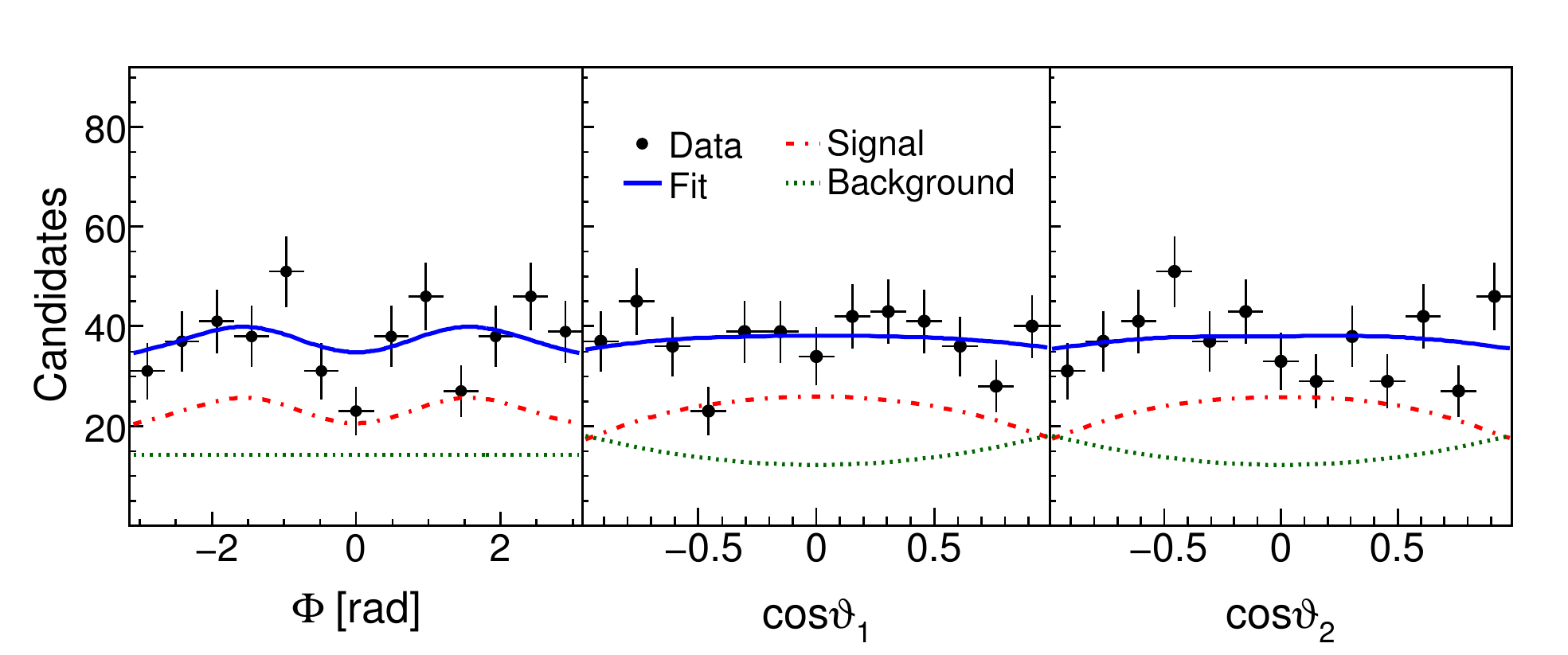}
\caption{\label{fig:projection}
Angular distribution for \phiphi\ 
events with the fit projection, signal, and background component superimposed.}
\end{figure}
The results of the polarization analysis for the \phiphi\ 
sample are summarized in Table~\ref{tab:summResults}. In
Fig.~\ref{fig:projection} we show the fit projections onto the
helicity angles. The dominant correlation of the fit
parameters is between $|A_0|^2$ and $|A_\parallel|^2$ (-0.447),
the others being much smaller.
%
\begin{table}[th]
\caption{\label{tab:summResults} 
Summary of the \phiphi\ measurements. The first uncertainty
quoted is statistical and the second is systematic.
\ifprlout
\else
\vspace {0.5cm}
\fi
}
\ifprlout
\begin{ruledtabular}
\else
\fi
\begin{tabular}{lc}
     Observable        &  Result \\
\hline
$\BR$                  &  $[2.32 \pm 0.18  \pm 0.82 ] \times 10 ^{-5} $ \\
$|A_0|^2$              & $0.348 \pm 0.041 \pm 0.021$ \\
$|A_\parallel|^2$      & $0.287 \pm 0.043 \pm 0.011$ \\
 $|A_\perp|^2$ & $0.365 \pm 0.044 \pm  0.027$   \\
$\cos\delta_\parallel$ & $-0.91^{+0.15}_{-0.13} \pm 0.09$    \\
\hline 
$A_u$              &  $-0.007 \pm 0.064 \pm 0.018$ \\
$A_v$              &  $-0.120 \pm 0.064 \pm 0.016$ \\
\end{tabular}
\ifprlout
\end{ruledtabular}
\else
\fi
\end{table}
%
%
Several sources of systematic uncertainty have been studied. 
We account for the neglected physics backgrounds
considering the \phikstpdg\ decay and two other possible
contaminations: $B^0_s \myto \phi
f_0(980)$, with $f_0\myto K^+K^-$, and $B^0_s \myto \phi K^+K^-$ (non--resonant). 
The latter two contributions are normalized to the signal yield in analogy with
similar $B^0\myto \phi X$ decays. We assume up to  4.6\% contamination from
$B^0_s\to\phi f_0$ and 0.9\% of $B_s^0\to \phi K^+K^-$, and determine a
1.5\%(0.4\%) shift in the central value for  $|A_0|^2$($|A_\parallel|^2$) using simulated experiments. 
Biases introduced by the time integration are
examined with MC simulation: they are created by the dependence of the angular acceptance on
$\Delta\Gamma_s$ and by a non--uniform acceptance in the $B^0_s$
proper decay time introduced by the displaced--track trigger. The assigned
systematic uncertainty (1\%) is the full shift expected in the central value,
assuming a value for $\Delta\Gamma_s$ equal to the world average plus
one standard deviation~\cite{pdg2010}. We also consider the propagation of $\tau_{\textup{L(H)}}$
uncertainties to the polarization amplitudes (1\%). 
Other sources of minor systematic uncertainties are the modeling
of the combinatorial background (0.4\%) and of the angular acceptance (0.5\%).
The impact of \CP--violating effects on the measured
amplitudes is negligible. 
\newline
\indent The asymmetries $A_i$ ($i=u,v$) are evaluated
through an unbinned ML fit to $m_B$ only, using 
the  joint likelihood for the $N^+_i$ and $N^-_i$ events with positive and negative $u$($v$).
The same $m_B$ PDF parameterization discussed above is used for samples with both $u$($v$) signs.
We multiply the total likelihood by the binomial 
$f(N^+_i, N^-_i|p)$,  where the probability $p$ of obtaining $N^+_i$ and $N^-_i$ events 
depends on the overall signal
fraction $f_s$, the signal asymmetry $A_i$, and the background
asymmetry $A_b^i$:
$p = \frac{1}{2}[1 + A_if_s+ (1-f_s)A_b^i]$.
%
Mass and width for the \Bs\ signal, as well as signal fraction, 
are consistent with those obtained in the polarization analysis,
while background asymmetries are consistent with zero.  
The measured \phiphi\ asymmetries are
reported in Table~\ref{tab:summResults}.
The systematic uncertainty is evaluated  using an alternate 
background parameterization as in the polarization analysis and by 
conservatively assigning maximal asymmetry to the neglected physics
background peaking in the signal region. 
Using a large sample of simulated events,
we check that the detector acceptance and resolution
introduce a bias in the asymmetries smaller than 0.2\%.
\newline
\indent 
 In summary, we measure for the first time the polarization amplitudes and 
the triple product asymmetries in the \phiphi\ decay.
We find a significantly suppressed longitudinal fraction
$f_{\text{L}}=|A_0|^2=0.348\pm0.041({\rm stat})\pm0.021({\rm syst})$, 
smaller than in other $b \myto s$ penguin \BVV\ decays~\cite{phikstar}. 
This result agrees well with predictions~\cite{BVV-QCDf} 
based on QCD factorization, but only marginally with perturbative QCD ones~\cite{Ali:2007ff},
and hints at a large penguin annihilation contribution~\cite{Datta:2008wf}.
The two measured asymmetries are statistically consistent with the no \CP\ violation
hypothesis, although  $A_v$
is $1.8\, \sigma$ different from zero.
\newline
\indent 
We thank D.~London, A.~Datta, M.~Gronau, and I.~Bigi for valuable discussions on time--integrated TP asymmetries.
We thank the Fermilab staff and the technical staffs of the participating institutions for their vital contributions. This work was supported by the U.S. Department of Energy and National Science Foundation; the Italian Istituto Nazionale di Fisica Nucleare; the Ministry of Education, Culture, Sports, Science and Technology of Japan; the Natural Sciences and Engineering Research Council of Canada; the National Science Council of the Republic of China; the Swiss National Science Foundation; the A.P. Sloan Foundation; the Bundesministerium f\"ur Bildung und Forschung, Germany; the Korean World Class University Program, the National Research Foundation of Korea; the Science and Technology Facilities Council and the Royal Society, UK; the Russian Foundation for Basic Research; the Ministerio de Ciencia e Innovaci\'{o}n, and Programa Consolider-Ingenio 2010, Spain; the Slovak R\&D Agency; the Academy of Finland; and the Australian Research Council (ARC).

%
%


\begin{thebibliography}{99}
\bibitem{CDFsystem}
CDF uses a cylindrical coordinate system with the $z$ axis along the proton beam axis. 
Pseudorapidity is $\eta  \equiv \ln(\tan(\vartheta/2))$, where $\vartheta$
is the polar angle, and $\phi$ is the azimuthal angle while $p_T = |p| \sin(\vartheta)$.
\bibitem{Lxy}
$\Lxy$ is calculated 
as the displacement of the  two-track intersection point with respect to the beam axis, 
projected onto  the total transverse momentum of the track pair.
\end{thebibliography}
\end{document}